\documentclass[prb,aps,twocolumn,amsmath,amssymb,floatfix]{revtex4-1}%
\usepackage[pdftex]{graphicx}
\usepackage{amsmath,amssymb,array,dcolumn,rotating,verbatim}
\usepackage{bm}

\begin{document}

\title{Consistent Asymptotic Expansion of Mott's Solution for Oxide Growth }

\author{Matthew R. Sears and Wayne M. Saslow}
\affiliation{Department of Physics, Texas A\&M University, \\
College Station, Texas 77840-4242 }

\begin{abstract}
Many relatively thick metal oxide films grow according to what is called the parabolic law $L=\sqrt{2A}t+\dots$.  Mott explained this for monovalent carriers by assuming that monovalent ions and electrons are the bulk charge carriers, and that their number fluxes vary as $t^{-1/2}$ at sufficiently long $t$.  In this theory no charge is present in the bulk, and surface charges were not discussed.  However, it can be analyzed in terms of a discharging capacitor, with the oxide surfaces as the plates.  The theory is inconsistent because the field decreases, corresponding to discharge, but there is no net current to cause discharge.  The present work, which also includes non-monovalent carriers, systematically extends the theory and obtains the discharge current.  Because the Planck-Nernst equations are nonlinear (although Gauss's Law and the continuity equations are linear) this leads to a systematic order-by-order expansion in powers of $t^{-1/2}$ for the number currents, concentrations, and electric field during oxide growth.  At higher order the bulk develops a non-zero charge density, with a corresponding non-uniform net current, and there are corrections to the electric field and the ion currents.  The second order correction to ion current implies a logarithmic term in the thickness of the oxide layer: $L=\sqrt{2A}t+B\ln{t}+\dots$.  It would be of interest to verify this result with high-precision measurements. 
\end{abstract}
\maketitle
\section{Introduction}

From the late 1930s to the late 1940s, N.F. Mott considered \cite{Mott1939a,Mott1940a,Mott1947a} the implications of experimental results \cite{Tammann1920} for oxide growth.  (For more of a review, see Ref~\onlinecite{Cabrera1949}.)  Under certain conditions (particularly high temperature), many metals develop \cite{Tammann1920} a layer of oxide at a parabolic rate on surfaces exposed to gas containing oxygen, according to 
\begin{equation}
L^2 = 2 A t, \label{eq:Thickness}
\end{equation}
where $L$ is the thickness of the oxide, $t$ is time, and $A$ is a constant.  The rate of growth is thus
\begin{equation}
\frac{dL}{dt} = \sqrt{\frac{A}{2}} t^{-1/2}. \label{eq:GrowthRate}
\end{equation}
This result may be thought of as representing the long-time asymptotic limit.  For specificity, we assume that the metal M fills $x<0$, that oxide MO fills $0<x<L$, and that oxygen gas O fills $L<x$.  This means we employ a moving coordinate system where $x=0$ represents the M/MO interface, and $x=L(t)$ represents the MO/O interface.  

A field and fluxes that vary as $t^{-1/2}$ are expected on the basis of a gradient of concentration, with the values of the carrier concentrations pinned by the two surfaces and the length $L$ determining the gradient.\cite{Tammann1920} That is, $dL/dt\sim 1/L$ gives a parabolic law.  Wagner obtained a parabolic growth law using the Planck-Nernst equations and some additional assumptions.\cite{Wagner1966}  Mott obtained a parabolic growth law using a more complete argument\cite{Mott1940a} that invokes the Planck-Nernst equations, Gauss's Law, and (implicitly) the continuity equations.  
In this case one can think of the electrochemical potentials pinned by the two surfaces and $L$ determining the gradient, which then yields the parabolic law.  

For electron and ion number currents ($j_{a}, j_{b}$) and ion valence $Z=1$, Mott assumed that the total current $J=e(j_b -j_a)$ in the oxide is zero, so
\begin{equation}
j_a = j_b. \label{eq:ZeroCurrent}
\end{equation}  
Since oxide grows when metal ions reach the oxide/gas interface, the growth rate is
\begin{equation}
\frac{dL}{dt} = j_{b} \Omega, \label{eq:CurrentDrivenGrowth}
\end{equation}
where $\Omega$ is the volume per metal ion in the newly formed oxide.  Comparison with (\ref{eq:GrowthRate}) immediately shows that 
\begin{equation}
j_{b} \sim t^{-1/2}\label{eq:j_Asympt}
\end{equation}
for the asymptotic behavior of the ion fluxes.  By the Planck-Nernst equations, the electric field $E$ and the ion density gradients $(\partial_{x}n_{a}, \partial_{x}n_{b})$ also have the same behavior.  Moreover, the quantities $(j_{a}, j_{b}, E, \partial_{x}n_{a}, \partial_{x}n_{b})$ are all uniform throughout the oxide. 

We assume that the metal and the gas are neutral, so by Gauss's Law the surface charges ($\Sigma(0), \Sigma(L)$) and the electric fields ($E(0), E(L)$) are related by
\begin{equation}
E(0)=\frac{\Sigma(0)}{\epsilon}, \qquad E(L)=-\frac{\Sigma(L)}{\epsilon}.\label{eq:ESig}
\end{equation}
Moreover, by continuity the assumption that there is charge and current only within the oxide leads to the conditions 
\begin{equation}
J(0)=\frac{d\Sigma(0)}{dt}, \qquad J(L)=-\frac{d\Sigma(L)}{dt}.\label{eq:JdSigdt}
\end{equation}

This model immediately poses the questions of whether Mott's solution is self-consistent, and whether it is the beginning of an asymptotic series in powers of $t^{-1/2}$.  To answer self-consistency, note that the uniform but decreasing-with-time $E\sim t^{-1/2}$ leads to no bulk charge and to interfaces with equal and opposite charge, so they behave like a capacitor.  Since $E$ decreases with time, so must the charge on the capacitor.  However, the model assumes zero current.  Hence Mott's solution is not self-consistent.  

Nevertheless we will show, in response to the question about asymptopia, that each of the continuous variables can be expanded in an asymptotic series in $t^{-n/2}$, where Mott's solution corresponds to $n=1$, and that non-zero current $J$ appears at order $n=3$.  We will also show that all of the continuous variables can depend upon position.  This means that the bulk can develop a local and total charge density, with the surfaces not having equal and opposite charges, so that the capacitor model holds only to lowest order.  It is already known that steady non-equilibrium current flow can cause local charge densities in the bulk.\cite{Saslow1996a, Yang1998a}

The fact that an expansion can be made of $j_{b}$ in powers of $t^{-1/2}$ leads to 
\begin{equation}
\frac{dL}{dt} = \sqrt{\frac{A}{2}} t^{-1/2}+B t^{-1} + \dots, \label{eq:GrowthRate2}
\end{equation}
so that  
\begin{equation}
L=\sqrt{2A}t^{1/2}+B\ln t +\dots
\end{equation}
to higher accuracy than given by the pure parabolic law.  This prediction can be subjected to  experimental study.  In practice, we must assume that 
\begin{equation}
L=\sqrt{2A}t^{1/2}+B\ln t + C +\dots
\end{equation}
because $\ln t$ is of order unity.  We have not found data of a high enough precision to verify this  $\ln{t}$ correction.

Section II gives the five fundamental equations for the five continuous variables (two continuity equations, Gauss's Law, and two Planck-Nernst equations), and indicates the expansion in powers of $t^{-1/2}$.  (The fact that it is such an expansion relates to the leading order term and the fact that the Planck-Nernst equation is a nonlinear function of the density.)  Section III gives the $n=1$ solution, which has two interface-reaction-rate-determined integration constants:  both $A$ and a constant-in-space density not included in the Mott solution.  Section IV presents the results of the $n=2$ solution that gives the first correction to Mott's solution, and shows that there is a non-zero bulk charge density and a corresponding spatial variation to the field.  There are four interface-reaction-rate-determined integration constants for $n=2$.  All higher-order solutions involve four interface-reaction-rate-determined integration constants.   Section V considers the oxide thickness growth rate.  Section VI provides a summary and conclusions.   
The Appendices explicitly give the $n=2$ and $n=3$ solutions.  
%

\section{On Two-Component Transport}
Let the subscripts $a$ and $b$ denote electrons and metal ions respectively.  Let $n_0$ and $n_0/Z$ be the uniform equilibrium concentration of electrons and metal ions respectively, ($n_a$,$n_b$) be their additional concentrations, ($\nu_a$,$\nu_b$) be their mobilities, ($D_a$,$D_b$) be their diffusion coefficients, ($q_a=-e{\rm{,}} \quad q_b=Z e$) be their charges, $E$ be the electric field, $k_B$ be the Boltzmann constant, $T$ be the temperature, and $x$ be the position.  

Note that we use the Einstein Relations to rewrite the mobilities, which can have either sign, in terms of diffusion constants, which are always positive.  Since we always consider $Z$ positive, if we want to consider oxygen ions and holes as the carriers, then only the sign of the electric charge $e$ must be changed.  For M$^{3+}$ and O$^{2-}$, we let $e \rightarrow 2e$ and $Z=3/2$.  Therefore our results are quite general.

\subsection{Equations for Two-Component Transport}
We will employ the Einstein Relations  
\begin{equation}
\frac{\nu_a}{D_a} = \frac{q_a}{k_B T} {\rm{,}} \quad \frac{\nu_b}{D_b} = \frac{q_b}{k_B T} \rm{.}\label{eq:Einstein}
\end{equation}
For electrons and metal ions, $- Z q_a = q_b$, so
\begin{equation}
\nu_b D_a = -Z \nu_a D_b , \quad - \frac{\nu_a}{D_a} = \frac{\nu_b}{Z D_b} = \frac{1}{V_T} ,\label{eq:Einstein2}
\end{equation}
where 
\begin{equation}
V_T = \frac{k_B T}{e} \label{eq:V_T}
\end{equation}
denotes a thermal voltage.
Mott and Cabrera's paper\cite{Cabrera1949} used ($v_a$,$v_b$) to denote electron and ion mobilities, which they may have intended to be ($\nu_{a},\nu_{b}$).

The one-dimensional Planck-Nernst equations for the number flux densities associated with metal ions and electrons are 
\begin{align}
&j_a = \nu_a (n_0 + n_a) E - D_a \partial_x n_a, \label{eq:PNaOriginalNu}\\
&j_b = \nu_b \left( \frac{n_0}{Z} + n_b \right) E - D_b \partial_x n_b. \label{eq:PNbOriginalNu}
\end{align}
Rewriting mobilities in terms of diffusion constants using \eqref{eq:Einstein},
\begin{align}
&j_a = -\frac{D_a}{V_T} (n_0 + n_a) E - D_a \partial_x n_a, \label{eq:PNaOriginalD}\\
&j_b = \frac{D_b}{V_T} (n_0 + Z n_b) E - D_b \partial_x n_b, \label{eq:PNbOriginalD}
\end{align}

We also use the number continuity equations,
\begin{equation}
\partial_t n_a + \partial_x j_a = 0, \quad 
\partial_t n_{b} + \partial_x j_{b} = 0 \label{eq:Continuityab} ,
\end{equation}
and Gauss's Law,
\begin{equation}
\partial_x E = \frac{e}{\epsilon}(Z n_b - n_a) \label{eq:Gauss}\rm{,}
\end{equation}
where $e$ is electron charge in Coulombs, and $\epsilon$ is the permittivity of the oxide.  

With the charge density and current density defined by
\begin{equation}
\rho=-e(n_{a}-Zn_{b}), \qquad J=-e(j_{a}-Zj_{b})
\label{eq:charge-current} ,
\end{equation}
use of the number continuity equations yields the charge continuity equation
\begin{equation}
\partial_t \rho + \partial_x J = 0. \label{eq:ChargeContinuity}
\end{equation}


\subsection{Expansion Notation}

We seek a series solution in powers of $t^{-1/2}$ for the electron and ion concentrations (densities) and fluxes, as well as for the electric field.  They must satisfy the Planck-Nernst equations, the continuity equations, and Gauss's Law.  
Following Mott, we take the lowest order fluxes and field to vary as $t^{-1/2}$.  Examining the structure of the Planck-Nernst equation for electrons \eqref{eq:PNaOriginalD}, and inserting terms of order $t^{-1/2}$, the nonlinear term $n_a E$ will contain terms of order $t^{-1}$.  Iteration yields that the series must be in powers of $t^{-n/2}$ for integer $n$.  


We thus make an expansion of the form
\begin{align}
&j_a = \sum_{n=1} J_{an} t^{-n/2}  ,  \quad
j_b = \sum_{n=1} J_{bn} t^{-n/2} \label{eq:jajbOriginal},\\
&n_a = \sum_{n=1} N_{an} t^{-n/2}  , \quad
n_b = \sum_{n=1} N_{bn} t^{-n/2} \label{eq:nanbOriginal},\\
&\Sigma^{(0)} = \sum_{n=1} \Sigma_n^{(0)} t^ {-n/2} ,\quad 
\Sigma^{(L)} = \sum_{n=1} \Sigma_n^{(L)} t^ {-n/2}\label{eq:Sigma0SigmaLOriginal}{\rm{,}}\\
&E=\sum_{n=1} E_{n} t^{-n/2} \label{eq:EOriginal},\\
&J=\sum_{n=1} J_{n} t^{-n/2}, \quad \rho = \sum_{n=1} \rho_{n} t^{-n/2}.
\end{align}
Here $J_{an}$, $J_{bn}$, $N_{an}$, $N_{bn}$, $E_n$, $\rho_n$, and $J_n$ are functions of the position along the direction of growth, $x$.  From the above definitions, the dimensionality of ($J_{an},J_{bn}$) is concentration times velocity times $t^{n/2}$, the dimensionality of ($N_{an},N_{bn}$) is concentration times $t^{n/2}$, the dimensionality of the surface charge density ($\Sigma_n^{(0)},\Sigma_n^{(L)} $) is charge per area times $t^{n/2}$, the dimensionality of $E_n$ is electric field times $t^{n/2}$, the dimensionality of $\rho_n$ is charge density times $t^{n/2}$, and the dimensionality of $J_n$ is current density times $t^{n/2}$. 

\subsection{On Specifying Chemical Reaction Rates at Surfaces}
In the presence of a true chemical reaction at a surface there is a single reaction rate, typically specified by a Butler-Volmer relation,\cite{Crow1994, Rubi2003a,Fleig2005} between the fluxes of all of the relevant components.  In the present case the fluxes of the carriers are independent of one another, so that there are two statements about carrier fluxes at each surface, for a total of four conditions.  With $\tilde\mu$ denoting an electrochemical potential, near equilibrium (as we have here, in the asymptotic regime) each flux $j$ will be proportional to its corresponding $\Delta\tilde\mu$ across the interface (either M/MO or MO/O).  The proportionality constant will depend details of the reaction and the baseline properties of the system.  That is, 
\begin{equation}
j_{a,b}=G_{a,b}\Delta\tilde\mu_{a,b}
\label{eq:BCs}
\end{equation} 
at each surface, so there are four $G$'s.  The equation becomes nonlinear far from equilibrium.  Thus the $j_{a,b}$ are proportional to a non-equilibrium quantity, which we take to be a field $E_{1}$, as in Ref.~\onlinecite{Cabrera1949}.  All of the unknown integration constants will be linear or higher in $E_{1}$.  

We will not attempt to carry this procedure any further.  It is sufficient for our purposes to know that this can be done, and that in the present problem there are four constants associated with boundary conditions at the two surfaces for the two carriers.  In principle, all of the quantities appearing in the solutions to the transport 
equations are determined by these surface reaction rates.  For a true chemical reaction, which we expect to be described by a Butler-Volmer equation, the fluxes at each surface, because they are related, will be described by only a single independent coefficient $G$.  Note also that the Butler-Volmer equation is non-linear, so that the boundary conditions can be nonlinear.  Because we do not consider the boundary conditions in detail, we will neglect this possibility.  

\section{Relations between expansion coefficients: all $n$}
\subsection{Continuity Relations and Charge Conservation}
The continuity equations imply charge conservation, so we treat them in the same subsection. 

The continuity equations \eqref{eq:Continuityab} yield
\begin{equation}
\begin{split}
\sum_{n=1} (\partial_x J_{an})t^{-n/2} + \sum_{m=1} N_{am} \left(- \frac{m}{2} \right) t^{-(m+2)/2}
 = 0 ,
 \end{split}
\end{equation}
for subscript $a$, and a similar relation holds for $b$.  With $m = (n-2)$, so that $\sum_{m=1} \rightarrow \sum_{n=3} $, comparison of like powers of $t$ yields, for $n=1$ and $n=2$,
\begin{equation}
\partial_x J_{an} = 0 , \quad \partial_x J_{bn} = 0  , \qquad \qquad \quad (n =1,2) ,\label{eq:PartialJa12Jb12}
\end{equation}
and, for $n \geq 3$,
\begin{align}
&\partial_x J_{an} = \left( \frac{n-2}{2} \right) N_{a(n-2)} , \notag\\
&\partial_x J_{bn} = \left( \frac{n-2}{2} \right) N_{b(n-2)}, \qquad \quad \qquad (n \geq 3) . \quad \label{eq:PartialJanJbn}
\end{align}
By definition we have 
\begin{equation}
\rho_{n}=e \left( ZN_{bn}-N_{an} \right), \qquad J_{n}=e \left(ZJ_{bn}-J_{an} \right), \label{eq:rho_n-J_n}
\end{equation}
and charge conservation for each $n$ is  
\begin{equation}
\partial_{t}\rho_{n}+\partial_{x} J_{n}=0.
\label{eq:ChCons_n}
\end{equation}

Charge conservation at each surface yields
\begin{equation}
\begin{split}
&\frac{d \Sigma^{(0)}}{dt} = -J |_{x=0} = -e (Zj_b - j_a)|_{x=0} , \\
&\frac{d \Sigma^{(L)}}{dt} = J |_{x=L} = e (Z j_b - j_a)|_{x=L} , \label{eq:SurfaceChargeCurrent}
\end{split}
\end{equation}
so
\begin{align}
&\sum_{n=1} \left(-\frac{n}{2}\right)\Sigma_n^{(0)} t^ {-(n+2)/2} \notag\\
&\qquad \quad = -e \sum_{n=1} \left(Z J_{bn}-J_{an}\right)|_{(x=0)} t^{-n/2} {\rm{,}} \\ &\sum_{n=1} \left(-\frac{n}{2}\right)\Sigma_n^{(L)} t^ {-(n+2)/2} \notag\\
&\qquad \quad = e \sum_{n=1} \left(Z J_{bn}-J_{an}\right)|_{(x=L)} t^{-n/2} \rm{.}
\end{align}
Comparing powers of $t$ we have, for $n=1$ and $n=2$,
\begin{align}
&Z J_{bn}|_{(x=0)}=J_{an}|_{(x=0)} ,\notag\\
& Z J_{bn}|_{(x=L)}=J_{an}|_{(x=L)}  ,\qquad \qquad (n =1,2) \rm{,}\label{eq:SigmaJ12}
\end{align}
and, for $n \geq 3$,
\begin{align}
&-\left(\frac{n-2}{2}\right)\Sigma_{(n-2)}^{(0)} = -e \left(Z J_{bn}-J_{an}\right)|_{(x=0)},   \notag\\
& -\left(\frac{n-2}{2}\right)\Sigma_{(n-2)}^{(L)} = e \left(Z J_{bn}-J_{an}\right)|_{(x=L)}, \notag\\
& \qquad \qquad \qquad \qquad \qquad \qquad \qquad \quad \quad \quad (n \geq 3) . \label{eq:SigmaJn}
\end{align}

%
Charge conservation over both surface and bulk yields
\begin{equation}
\begin{split}
&\sum_{n=1} \left( \Sigma_{n}^{(0)} + \Sigma_{n}^{(L)} \right) t^{-n/2} \\
&\qquad \quad =  e \int_0^L \sum_{n=1}(N_{an} - Z N_{bn})t^{-n/2}dx, 
\end{split}
\end{equation}
so that, for each $n$,
\begin{equation}
\Sigma_{n}^{(0)} + \Sigma_{n}^{(L)} = e \int_0^L (N_{an} - Z N_{bn}) dx \rm{.} \label{eq:SigmaTerms}
\end{equation}

\subsection{Gauss's Law}
Gauss's Law \eqref{eq:Gauss} reads
\begin{equation}
\sum_{n=1} (\partial_x E_n) t^{-n/2} = \frac{1}{\epsilon} \left( \sum_{n=1} \rho_{n} t^{-n/2} \right),
\end{equation}
so, for each $n$,
\begin{equation}
\partial_x E_n = \frac{1}{\epsilon} \rho_{n}. \label{eq:GaussTerms}
\end{equation}
Gauss's Law at the surfaces \eqref{eq:ESig} gives, for each n,
\begin{equation}
E_n(0) = \frac{\Sigma_{n}^{(0)}}{\epsilon} {\rm{,}} \quad E_n(L) = -\frac{\Sigma_{n}^{(L)}}{\epsilon} \rm{.} \label{eq:GaussSurfaceTerms}
\end{equation}

\subsection{Planck-Nernst}
The Planck-Nernst equation for species $a$, \eqref{eq:PNaOriginalD}, can be written as
\begin{equation}
\begin{split}
&\sum_{n=1} J_{an} t^{-n/2} = -\frac{D_a}{V_T} n_{0}\sum_{n=1} E_n t^{-n/2}\\
& \quad -\frac{D_a}{V_T} \sum_{m,n=1} N_{an} E_m t^{-(m+n)/2} - D_a \sum_{n=1} (\partial_x N_a) t^{-n/2} \rm{.}
\end{split}
\end{equation}
or,
\begin{equation}
\begin{split}
&\sum_{n=1} (J_{an} + \frac{D_a}{V_T} n_{0}E_n + D_a \partial_x N_{an}) t^{-n/2} \\
&\qquad \quad = -\frac{D_a}{V_T} \sum_{m,n=1} N_{an} E_m t^{-(m+n)/2} \label{eq:FullPNa},
\end{split}
\end{equation}
with a similar form for species $b$,
\begin{equation}
\begin{split}
&\sum_{n=1} (J_{bn} - \frac{D_b}{V_T} n_{0}E_n + D_b \partial_x N_{bn}) t^{-n/2} \\
&\qquad \quad = \frac{Z D_b}{V_T} \sum_{m,n=1} N_{bn} E_m t^{-(m+n)/2} \label{eq:FullPNb}.
\end{split}
\end{equation}
By matching coefficients of powers of $t$, we obtain the resultant equations for any $n$.

\subsection{Solving the Transport Equations}
There are five first-order differential equations for five continuous variables, so there are five integration constants at each order.  For $n\ge3$ the current density $J_{n}$ at $x=L$ is known from the $n-2$ value of surface charge density $\Sigma^{(L)}_{n-2}$.  Therefore only four integration constants need be determined.  These can be thought of as fixed by the ``reaction rates'' of each charge carrier at each of the interfaces, which will not be specified. 

The cases $n=1$ and $n=2$ are a bit simpler than $n \geq 3$.  Nevertheless, for each $n$ our strategy will be the same: (1) from the continuity equations find the ion fluxes $J_{an}$ and $J_{bn}$; (2) from all of the five equations find an equation for $\rho_{n}$ and solve it; (3) use this $\rho_{n}$ in Gauss's Law to find $E_{n}$; (4) find $N_{an}$ and $N_{bn}$ by substitution of $J_{an}$, $J_{bn}$ and $E_n$ into the Planck-Nernst equations.

   
\section{Solution for $n=1$}
For $n=1$, the calculations are simple, but illustrate what happens in higher orders.  The continuity equations \eqref{eq:PartialJa12Jb12} give $J_{a1}$ and $J_{b1}$ to be uniform, and charge conservation at each surface \eqref{eq:SigmaJ12} gives
\begin{equation}
J_{a1} = Z J_{b1} \label{eq:Ja1Jb1equal}.
\end{equation}

The Planck-Nernst equations \eqref{eq:FullPNa} and \eqref{eq:FullPNb} yield
\begin{align}
&J_{a1} + \frac{D_a}{V_T} n_{0}E_1 + D_a \partial_x N_{a1} = 0, \label{eq:PNa1}\\
&J_{b1} - \frac{D_b}{V_T} n_{0}E_1 + D_b \partial_x N_{b1} = 0, \label{eq:PNb1}
\end{align}
where we take $E_1$ as a uniform, experimentally determined value.  
 
From the uniformity of $E_1$,
\begin{equation}
\partial_x E_1 =0 \rm{,}
\end{equation}
so that Gauss's Law yields
\begin{equation}
\rho_1 = 0 {\rm{,}} \quad N_{a1} = Z N_{b1} \rm{.} \label{eq:Na1Nb1Equal}
\end{equation}

Substitution of fluxes from \eqref{eq:Ja1Jb1equal} and concentrations from \eqref{eq:Na1Nb1Equal} into the Planck-Nernst equations gives
\begin{align}
&J_{a1} + \frac{D_a}{V_T}  n_{0}E_1 + D_a \partial_x N_{a1} = 0 \rm{,} \\
&\frac{J_{a1}}{Z} -\frac{D_b}{V_T} n_{0}E_1 + D_b \partial_x N_{b1} = 0 \rm{.} 
\end{align}
Thus, with \eqref{eq:Na1Nb1Equal}, we have

\begin{align}
&J_{a1} = Z J_{b1} = -(1+Z)  \frac{D_a D_b}{D_b - D_a} \frac{n_{0}E_1}{V_T} , \label{eq:Ja1Jb1Solved}\\
&N_{a1} = Z N_{b1} = \frac{H n_{0}E_1}{V_T} x + M_1 ; \label{eq:Na1Nb1Solved}
\end{align}
here,
\begin{equation}
H = \left( \frac{D_a + Z D_b}{D_b - D_a} \right)
\label{eq:H}
\end{equation}
is dimensionless,
and $M_1$ is a constant of integration (units of concentration times s$^{1/2}$) determined by the surface reaction rates (and thus linear in $E_{1}$).  (Recall that $(J_{a1},N_{a1})$ are first order coefficients, which must be multiplied by $t^{-1/2}$ in order to find the flux and number densities $(j,n)$.)  Although constants of integration associated with reaction rates were discussed earlier, we consider them more explicitly here. 

\begin{figure}[t]
\centerline{\includegraphics[width=6.7cm]{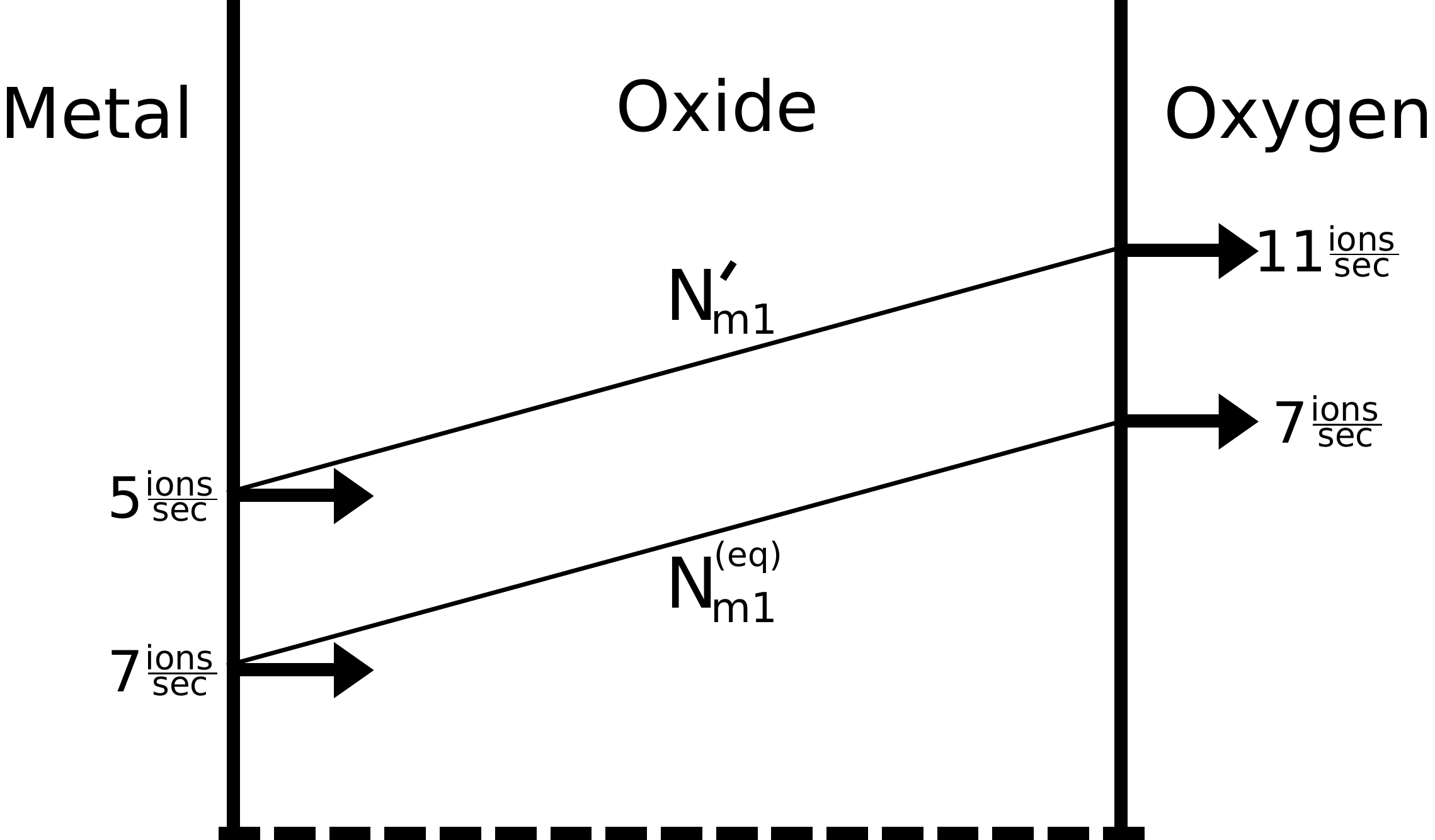}}
\caption{The effect of the constant concentration $M_1$ of metal ions.  Here $N_{m1}'$ has too high a value of $M_1$.  Only the particular value of $M_1$ in $N_{m1}^{(eq)}$ permits an equal rate of ions to enter and leave the oxide, here taken to be 7 ions per second.}
\label{fig:M1Graph}
\end{figure} 
Figure~\ref{fig:M1Graph} illustrates the effect of changing the value of $M_1$.
If $M_1$ were very large, the high concentration of metal ions near the metal/oxide surface would oppose new ions from entering, whereas the high concentration of metal ions near the oxide/gas interface would encourage more ions to be deposited on the oxide/gas surface.  Eventually, the number of metal ions in the bulk would be insufficient to maintain the high rate of ions exiting the oxide, and would drop to some equilibrium value.  Thus, $M_1$ is determined by constraining the oxide to have no net ion-loading or ion-unloading in the bulk at order $n=1$.  Note that $E_{1}$, which is proportional to (the parabolic growth rate coefficient), is also related to the surface reaction rates.  In general, net surface reaction rates involve a Butler-Volmer equation, but not far from equilibrium (as in the Mott solution) they can be linearized in the differences of various electrochemical potentials.  This will ensure that there is no net surface reaction rate in the limit of equilibrium.  


%
%
From Gauss's Law at the surfaces \eqref{eq:GaussSurfaceTerms},
\begin{equation}
E_1(0) = \frac{\Sigma_1^{(0)}}{\epsilon} {\rm{,}} \quad E_1(L) = -\frac{\Sigma_1^{(L)}}{\epsilon} \rm{.}
\end{equation}
Since $E_1$ is uniform, 
\begin{equation}
\Sigma_1^{(0)} = -\Sigma_1^{(L)} = \epsilon E_1 \rm{.} 
\label{eq:Sigma1}
\end{equation}

For $|\nu_a| >> |\nu_b|$ (or equivalently in this case, $D_a >> D_b$), Mott and Cabrera \cite{Cabrera1949} find for monovalent ions that 
\begin{equation}
J_{a1} = -2 D_b \frac{\partial N_{a1}}{\partial x}; 
\label{lim Ja1}
\end{equation} 
our results can be shown to be consistent with this. 


\section{Solution for $n=2$}
We here summarize the $n=2$ results.  (For the explicit solution, see Appendix \ref{sec:n2appendix}.)  Recall that all coefficients must be multiplied by $t$ to find the physical variables $(j,n,E)$.  With the constant $M_{21}$ (in units  s$/$m$^2$) determined by surface reaction rates (and thus linear in $E_{1}$), the second order flux density coefficients are
\begin{align}
  J_{a2} = Z J_{b2} = - (1+Z) \left( \frac{D_a D_b}{Z D_b + D_a} \right) M_{21}, \label{eq:Ja2Jb2Summary}
 \end{align}
and there is no net charge flux,
\begin{equation}
J_2 =  ZeJ_{b2}-eJ_{a2} = 0.
\end{equation}  

With the constants $M_{20}$, $P_{a2}^{(+)}$ and $P_{a2}^{(-)}$ (each in units of s/m) determined by surface reaction rates (and thus linear in $E_{1}$), the $n=2$ coefficients of the concentrations of electrons and ions are given by
\begin{align}
&N_{a2} = M_{20} + M_{21}x   + P_{a2}^{(+)}e^{x/l_s} + P_{a2}^{(-)}e^{-x/l_s} ,\\
&N_{b2} =    \frac{M_{20}}{Z} - \frac{1}{Z}  \frac{H \epsilon E_1^2}{V_T e} + \frac{M_{21}}{Z}x\notag\\
&\qquad \qquad \qquad - P_{a2}^{(+)}e^{x/l_s} - P_{a2}^{(-)}e^{-x/l_s} . 
\end{align}
 Here, $l_s$ is the screening length,
 \begin{equation}
 l_s = \sqrt{\frac{V_T \epsilon}{(1+Z) n_{0}e}} = \sqrt{\frac{k_B T \epsilon}{(1+Z) n_{0}e^2}}.
 \end{equation}
 There is a net charge in the bulk, given by $ \rho_2 t^{-1}$, where
 \begin{equation}
 \begin{split}
 &\rho_2 =  P_2^{(+)} e^{x/l_s} + P_2^{(-)} e^{-x/l_s}  -  \frac{H \epsilon E_1^2}{V_T}  ,
 \end{split}
 \end{equation}
and
 \begin{equation}
 P_2^{(+)} = - (1+Z) e P_{a2}^{(+)}, \quad P_2^{(-)} = - (1+Z) e P_{a2}^{(-)}.
 \end{equation}
Note that $\rho_2$ has, in addition to surface charge within a screening length of the two surfaces, a uniform charge density with sign determined by $e/(D_a-D_b)$ and independent of the sign of $E_1$ (or, equivalently, the direction of current flow).  Since $D_a \gg D_b$ here, the term is positive.  For holes and oxygen ions the carriers, we have $-e/(D_a - D_b)$, but $D_b \gg D_a$, so it is again positive.  As found in previous work,\cite{Saslow1996a} this uniform charge density leads to a quadratic voltage profile within the bulk, not within a screening length of either surface.
 
 The second order coefficient of the electric field is
 \begin{align}
&E_2 =  \frac{ l_s}{\epsilon}\left( P_2^{(+)} e^{x/l_s} - P_2^{(-)} e^{-x/l_s} \right) -  \frac{H E_1^2}{V_T} x \notag\\
&\qquad + \frac{V_T M_{21}}{n_{0} H}  - \frac{M_1 E_1}{n_{0}}. 
\end{align}

The surface charge coefficients are given by
\begin{align}
&\Sigma_2^{(0)} =  l_s\left( P_2^{(+)} - P_2^{(-)} \right)+  \frac{\epsilon V_T M_{21}}{n_{0} H}  - \frac{\epsilon M_1 E_1}{n_{0}},  \\
&\Sigma_2^{(L)} =  - l_s\left( P_2^{(+)} e^{L/l_s} - P_2^{(-)} e^{-L/l_s} \right) +  \frac{\epsilon H E_1^2}{V_T} L \notag\\
&\qquad  -  \frac{\epsilon V_T M_{21}}{n_{0} H}  + \frac{\epsilon M_1 E_1}{n_{0}}. 
\end{align}
We have verified that
\begin{equation}
\Sigma_2^{(0)} + \Sigma_2^{(L)} + \int_0^L \rho_2 dx =0,
\end{equation}
so there is no net charge in the system.

\section{Rate of Growth of Oxide Layer}
The oxide layer grows as metal ions reach the MO/O surface, and are taken into lattice positions to form a new oxide layer.  Thus, the rate of growth of the oxide depends on the rate $j_{b}$ at which metal ions arrive, according to \eqref{eq:CurrentDrivenGrowth}, or $dL/dt=\Omega j_{b}$.  
Including $n=3$ (see Appendix \ref{sec:n3appendix}) we find the metal ion number flux (using Eqs. \eqref{eq:Ja1Jb1Solved}, \eqref{eq:Ja2Jb2Summary} or \eqref{eq:Ja2Jb2Solved}, and \eqref{eq:Jb3}),
\begin{equation}
\begin{split}
j_b =& \left(-\left(\frac{1+Z}{Z}\right) \frac{D_a D_b}{D_b - D_a} \frac{n_{0}E_1}{V_T} \right) t^{-1/2}\\
&\quad  + \left( - \left(\frac{1+Z}{Z}\right) \left( \frac{D_a D_b}{Z D_b + D_a} \right) M_{21}\right) t^{-1}\\
&\quad + \left(\frac{H n_0 E_1}{4 V_{T} Z} x^2 + \frac{M_1}{2Z} x + \frac{K_{3}}{Z} + \frac{\epsilon E_1}{2 Z e} \right) t^{-3/2} + \dots \rm{,} \label{eq:Fulljb} 
\end{split}
\end{equation}
where $K_{3}$ is a constant of integration (units of flux density times s$^{3/2}$) determined by interfacial reaction rates (and thus linear in $E_{1}$).  Note that, if $D_{b}<D_{a}$ (as for ions relative to electrons) then the microscopics must give $E_{1}>0$ for a positive growth rate. Keeping only terms of second order, integration of \eqref{eq:Fulljb} with respect to time gives 
\begin{equation}
L = \sqrt{2A} t^{1/2} + B \ln{t} + \dots \rm{,}
\end{equation}
where 
\begin{align}
&A = 2 \left(\frac{1+Z^2}{Z} \right)^2  \left(\frac{D_a D_b}{D_b - D_a}\right)^2 \frac{n_{0}^2 E_1^2 \Omega^2}{V_T^2},\\
&B =  \frac{D_a D_b}{D_a + Z^2 D_b} \left( (1-Z) \frac{ M_1 E_1}{V_T} -\frac{1+Z^2}{Z}  M_{21} \right) \Omega \rm{.}
\end{align}


\section{Summary and Conclusion}
We have shown that the approach taken by Mott for parabolic growth of oxide films can be turned into a consistent asymptotic expansion, and we have explicitly given the form of the lowest three orders.  Up to four integration constants appear at each order, related to the surface reaction rates.  At higher order the bulk film is found to be charged, with a corresponding non-uniform current density.  

The Appendices present the $n=2$ and $n=3$ solutions in detail, to show that the method can be used for any $n$ to find the fluxes, concentrations, surface charges and electric field.  As a consequence one can have confidence that the Mott solution gives the leading term in the complete solution of the complete set of transport equations. 

The most easily verifiable prediction from the viewpoint of experiment is the prediction that the first correction to the linear growth law is logarithmic.  Because $\ln{t}$ is of order unity, data should be analyzed with an additional constant: $L=\sqrt{2A}t+B\ln{t}+C+\dots$.  A sampling of the current literature\cite{Dravnieks1951, Martin1994a, Lahiri1998, Jeurgens2002, Zhong2008a} did not find enough precision to confirm the logarithmic form of the correction term.

We would like to thank Allan Jacobson for his comments and suggestions.  This work was partially supported by the Department of Energy through grant DE-FG02-06ER46278.

\appendix
\section{Explicit Solution for $n=2$}
\label{sec:n2appendix}
By the continuity equations \eqref{eq:PartialJa12Jb12} $J_{a2}$ and $J_{b2}$ must be uniform,
and by \eqref{eq:SigmaJ12} they are related by
\begin{equation}
J_{a2} = Z J_{b2} \label{eq:Ja2Jb2Relation} .
\end{equation}

%
For $n=2$, the Planck-Nernst equations from \eqref{eq:FullPNa} and \eqref{eq:FullPNb} are
\begin{align}
&J_{a2} + \frac{D_a}{V_T} n_{0}E_2 + D_a \partial_x N_{a2} = -\frac{D_a}{V_T} N_{a1}E_{1}\rm{,} \label{eq:PNa2}\\
&J_{b2} - \frac{D_b}{V_T} n_{0}E_2 + D_b \partial_x N_{b2} = \frac{Z D_b}{V_T} N_{b1}E_{1}\rm{.} \label{eq:PNb2}
\end{align}
Taking spatial derivatives of \eqref{eq:PNa2} and \eqref{eq:PNb2}, and using Gauss's Law \eqref{eq:GaussTerms} and the uniformity of $J_{a2}$ and $J_{b2}$, yields 
\begin{align}
&\frac{D_a n_{0}}{V_T \epsilon}\rho_{2} + D_a \partial_x^2 N_{a2} = -\frac{D_a}{V_T} \partial_x (N_{a1}E_{1}),\\
&- \frac{D_b n_{0}}{V_T \epsilon}\rho_{2} + D_b \partial_x^2 N_{b2} = \frac{Z D_b}{V_T} \partial_x (N_{b1}E_{1}).
\end{align}
The right hand sides are found from \eqref{eq:Na1Nb1Solved};
\begin{align}
&\frac{D_a n_{0}}{V_T \epsilon}\rho_{2} + D_a \partial_x^2 N_{a2} = -D_a  \frac{H n_{0}E_1^2 }{V_T^2}\rm{,} \label{eq:PartialPNa2}\\
&- \frac{D_b n_{0}}{V_T \epsilon}\rho_{2} + D_b \partial_x^2 N_{b2} = D_b  \frac{H n_{0}E_1^2}{V_T^2} , \label{eq:PartialPNb2}
\end{align}
where $H$ is defined in \eqref{eq:H}.

Subtracting \eqref{eq:PartialPNa2} multiplied by $1/D_a$ from \eqref{eq:PartialPNb2} multiplied by $Z/ D_b$ yields
\begin{equation}
\begin{split}
\partial_x^2 \rho_2 - (1+Z) \frac{n_0 e}{V_T \epsilon} \rho_2 = (1+Z) \frac{H n_0 e E_1^2}{V_T^2}.
\label{eq:PartialDelta2}
\end{split}
\end{equation}
The solution to this equation, with 
\begin{equation}
l_s \equiv \sqrt{\frac{V_T \epsilon}{(1+Z) n_{0}e}} = \sqrt{\frac{k_B T \epsilon}{(1+Z) n_{0}e^2}}, \label{eq:ScreeningLength}
\end{equation}
and with new integration constants $P_2^{(+)}$ and $P_2^{(-)}$, is
\begin{equation}
\begin{split}
\rho_2 = P_2^{(+)} e^{x/l_s} + P_2^{(-)}  e^{-x/l_s}  -  \frac{H \epsilon E_1^2}{V_T} . \label{eq:Delta2Solved}
\end{split}
\end{equation}
From \eqref{eq:Delta2Solved} we infer that $N_{a2}$ and $Z N_{b2}$ are polynomials whose terms that are linear or higher are equal, and they may have different exponential terms.

Substituting $\rho_2$ from \eqref{eq:Delta2Solved} into Gauss's Law \eqref{eq:GaussTerms}, and integrating $\partial_x E_{2}$ yields 
\begin{equation}
\begin{split}
&E_2 = \frac{l_s}{\epsilon}\left( P_2^{(+)} e^{x/l_s} - P_2^{(-)} e^{-x/l_s} \right) -  \frac{H E_1^2}{V_T} x + F_2, \label{eq:E2Solved}
\end{split}
\end{equation}
where $F_2$ is a new integration constant with units V-s$/$m. 

By \eqref{eq:Na1Nb1Solved}, $(N_{a1},N_{b1})$ are linear in $x$, so that the right-hand-side of the Planck-Nernst equation \eqref{eq:PNa2} is linear in $x$.  Moreover, the continuity equation \eqref{eq:PartialJa12Jb12} implies that $J_{a2}$ is constant.  Therefore by \eqref{eq:E2Solved} for $E_2$, the Planck-Nernst equation allows $\partial_x N_{a2}$ to be linear, so $N_{a2}$ can be quadratic in $x$.  Moreover, the exponential terms in $(D_{a}/V_T)n_{0}E_2+D_{a}\partial_x N_{a2}$ must cancel.  Finally, from \eqref{eq:Delta2Solved} any linear or quadratic terms in $N_{a2}$ and $Z N_{b2}$ must be equal.  With ($M_{20},M_{21},M_{22},P_{a2}^{(\pm)}$) being new integration constants, we therefore conclude that the following form must hold:
\begin{equation}
N_{a2} = M_{20} + M_{21}x + M_{22} x^2  + P_{a2}^{(+)}e^{x/l_s} + P_{a2}^{(-)}e^{-x/l_s} . 
\label{eq:Na2Generic}
\end{equation}
Use of \eqref{eq:rho_n-J_n} for $n=2$ ($\rho_2 = e(Z N_{b2} - N_{a2})$), and $\rho_2$ of \eqref{eq:Delta2Solved} gives
\begin{align}
&N_{b2} = \frac{M_{20}}{Z} - \frac{1}{Z}  \frac{H \epsilon E_1^2}{V_T e} + \frac{M_{21}}{Z}x + \frac{M_{22}}{Z} x^2 \notag\\
&\quad + \left( \frac{P_{2}^{(+)} + e P_{a2}^{(+)}}{Z e} \right) e^{x/l_s} + \left( \frac{P_{2}^{(-)} + e P_{a2}^{(-)}}{Z e} \right) e^{-x/l_s} .
\label{eq:Nb2Generic}
\end{align}

Addition of \eqref{eq:PNa2} divided by $D_a$ and \eqref{eq:PNb2} divided by $D_b$ gives, with \eqref{eq:Ja2Jb2Relation} and \eqref{eq:Na1Nb1Solved},
\begin{equation}
\begin{split}
 & J_{a2} \left(\frac{1}{D_a} + \frac{1}{Z D_b} \right) + \partial_x \left(N_{a2} + N_{b2} \right) =0.
 \end{split}
\end{equation}
Substitution for $N_{a2}$ and $N_{b2}$ from \eqref{eq:Na2Generic} and \eqref{eq:Nb2Generic} allows us to solve for some of the constants,
\begin{align}
& M_{21} \left( 1 + \frac{1}{Z} \right) + 2 M_{22} \left(1 + \frac{1}{Z} \right) x  \notag\\
&+ \frac{1}{l_s} \left( P_{a2}^{(+)} \left(1 + \frac{1}{Z} \right) + \frac{ P_{2}^{(+)}}{Z e} \right) e^{x/l_s} \notag\\
&  -\frac{1}{l_s} \left( P_{a2}^{(-)}\left(1 + \frac{1}{Z} \right) + \frac{ P_{2}^{(-)}}{Z e} \right) e^{-x/l_s} \notag\\
& \qquad \qquad =-  J_{a2} \left(\frac{1}{D_a} + \frac{1}{Z^2 D_b} \right) .
\end{align}
So, since $J_{a2}$ is uniform, comparison of powers of $x$ yields the conditions
\begin{align}
&P_{2}^{(+)} = -(1+Z) e P_{a2}^{(+)}, \quad P_{2}^{(-)} = -(1+Z) e P_{a2}^{(-)},\\
&J_{a2} = Z J_{b2} = - (1+Z) \left( \frac{D_a D_b}{Z D_b + D_a} \right) M_{21} ,\label{eq:Ja2Jb2Solved}\\ 
&M_{22} = 0. \label{eq:M22} 
\end{align}
We may thus rewrite \eqref{eq:Na2Generic} and \eqref{eq:Nb2Generic} as
\begin{align}
&N_{a2} = M_{20} + M_{21}x   + P_{a2}^{(+)}e^{x/l_s} + P_{a2}^{(-)}e^{-x/l_s} , 
\label{eq:Na2}\\
&N_{b2} =    \frac{M_{20}}{Z} - \frac{1}{Z}  \frac{H \epsilon E_1^2}{V_T e} + \frac{M_{21}}{Z}x \notag\\
&\qquad \qquad \qquad - P_{a2}^{(+)}e^{x/l_s} - P_{a2}^{(-)}e^{-x/l_s} . \label{eq:Nb2}
\end{align}

Note that in \eqref{eq:PNa2} the exponential and linear coefficients already match, by construction.  A new constraint, however, is found by comparing the constant terms,
%
\begin{align}
& -  (1 + Z) \left( \frac{D_a D_b}{Z^2 D_b + D_a} \right) M_{21} + \frac{D_a n_{0}}{V_T} F_2  + D_a M_{21} \notag\\
&\qquad \qquad \qquad =  - \frac{D_a M_1 E_1}{V_T},
\end{align}
yielding the condition that
\begin{equation}
F_2 = \frac{V_T M_{21}}{n_{0} H}  - \frac{M_1 E_1}{n_{0}}. \label{eq:F2Solved}
\end{equation}

We are thus left with
four independent  constants of integration, which we take to be $M_{20}$, $M_{21}$, $P_{2}^{(+)}$, and $P_{2}^{(-)}$.  The reaction rates for electrons and ions at each interface, needed to produce the correct ion fluxes, provide the 4 conditions necessary to solve for these constants.

For completeness we note that, from Gauss's Law at each surface
\eqref{eq:GaussSurfaceTerms},
\begin{equation}
E_2(0) = \frac{\Sigma_2^{(0)}}{\epsilon} {\rm{,}} \quad E_2(L) = -\frac{\Sigma_2^{(L)}}{\epsilon} \rm{.}
\end{equation}
Then \eqref{eq:E2Solved} and \eqref{eq:F2Solved} yield
\begin{align}
&\Sigma_2^{(0)} =  l_s\left( P_2^{(+)} - P_2^{(-)} \right)+ \frac{\epsilon  V_T M_{21}}{n_{0} H}  - \frac{\epsilon M_1 E_1}{n_{0}} , \\
&\Sigma_2^{(L)} =  - l_s\left( P_2^{(+)} e^{L/l_s} - P_2^{(-)} e^{-L/l_s} \right) +  \frac{\epsilon H E_1^2}{V_T} L \notag\\
&\qquad - \frac{\epsilon  V_T M_{21}}{n_{0} H}  + \frac{\epsilon M_1 E_1}{n_{0}}  .
\end{align}

\section{Solution for $n=3$}
\label{sec:n3appendix}
Recall that in order $n=3$, all coefficients must be multiplied by $t^{3/2}$ to find the physical variables ($j,n,E$).


Using the $n=1$ coefficients for electron and metal ion number densities from \eqref{eq:Na1Nb1Solved} and the continuity equations from \eqref{eq:PartialJanJbn} for $n=3$,
\begin{align}
&\partial_x J_{a3} = \frac{H n_0 E_1}{2 V_T} x + \frac{M_1}{2} , \quad \partial_x J_{b3} = \frac{Hn_0 E_1}{2 V_{T} Z} x + \frac{M_1}{2Z} x .
\label{eq:PartialJ3}
\end{align}
Integration of these two equations gives two integration constants.  However, these two constatnts are constrained by charge conservation across the interface at $x=0$ \eqref{eq:SigmaJn}, 
\begin{equation}
Z J_{b3}|_{x=0} = J_{a3}|_{x=0} + \frac{\Sigma_1^{(0)}}{2 e},
\label{eq:SurfaceConservation3}
\end{equation}
where $\Sigma_1^{(0)}$ is given by \eqref{eq:Sigma1}.  
Integration of the continuity equations \eqref{eq:PartialJ3} then yields only a single new integration constant, which we call $K_{3}$: 
\begin{align}
&J_{a3} = \frac{H n_0 E_1}{4 V_T} x^2 + \frac{M_1}{2} x + K_{3}, \label{eq:Ja3}\\
&J_{b3} = \frac{Hn_0 E_1}{4 V_{T} Z} x^2 + \frac{M_1}{2Z} x + \frac{K_{3}}{Z} + \frac{\epsilon E_1}{2 Z e}. \label{eq:Jb3}
\end{align}
There is a uniform net electric charge flux (i.e., current density) at order $n=3$,
\begin{equation}
J_3 = e \left( ZJ_{b3} - J_{a3} \right) = \frac{\epsilon E_1}{2},
\end{equation}
due to the discharge of the surfaces in order $n=1$.

For $n=3$, the Planck-Nernst equations from \eqref{eq:FullPNa} and \eqref{eq:FullPNb} are 
\begin{align}
J_{a3} + \frac{D_a}{V_T} n_0 E_3 + D_a \partial_x N_{a3} = - \frac{D_a}{V_T} \left(N_{a2} E_1 + N_{a1} E_2 \right),\label{eq:PNa3}\\
J_{b3} - \frac{D_b}{V_T} n_0 E_3 + D_b \partial_x N_{b3} =  \frac{Z D_b}{V_T} \left(N_{b2} E_1 + N_{b1} E_2 \right).\label{eq:PNb3}
\end{align}
Taking spatial derivatives of \eqref{eq:PNa3} and \eqref{eq:PNb3}, and using Gauss's Law \eqref{eq:GaussTerms},
\begin{align}
&\partial_x J_{a3} + \frac{D_a n_0}{V_T \epsilon} \rho_3 + D_a \partial_x^2 N_{a3} \notag\\
&\qquad \qquad \qquad \qquad = - \frac{D_a}{V_T} \partial_x \left(N_{a2} E_1 + N_{a1} E_2 \right),\label{eq:PartialPNa3}\\
&\partial_x J_{b3} - \frac{D_b n_0}{V_T \epsilon} \rho_3 + D_b \partial_x^2 N_{b3} \notag\\
&\qquad \qquad \qquad  \qquad =  \frac{Z D_b}{V_T} \partial_x \left(N_{b2} E_1 + N_{b1} E_2 \right).\label{eq:PartialPNb3}
\end{align}
Subtracting \eqref{eq:PartialPNa3} multiplied by $1/D_a$ from \eqref{eq:PartialPNb3} multiplied by $Z/D_b$ yields an equation for $\rho_3$,
\begin{align}
&e \left(\frac{1}{D_b} - \frac{1}{D_a}\right) \partial_x J_{a3} - (1+Z)\frac{n_0 e}{V_T \epsilon} \rho_3 + \partial_x^2 \rho_3 =\notag\\
&\quad \frac{e E_1}{V_T} (\partial_x N_{a2} + Z^2 \partial_x N_{b2})  
 +\frac{e}{V_T} (\partial_x N_{a1} + Z^2 \partial_x N_{b1})E_2 \notag\\
&\quad +\frac{e}{V_T} (N_{a1} + Z^2 N_{b1})\partial_x E_2.
\end{align}
Substitution for $(N_{a1},N_{b1})$ from \eqref{eq:Na1Nb1Solved}, $(N_{a2},N_{b2})$ from \eqref{eq:Na2} and \eqref{eq:Nb2}, $E_2$ from \eqref{eq:E2Solved}, and $J_{a3}$ from \eqref{eq:Ja3} yields
\begin{widetext}
\begin{align}
&\partial_x^2 \rho_3 - \frac{1}{l_s^2} \Delta \rho_3 = \notag\\
& \qquad e \left(\frac{D_b - D_a}{D_a D_b}\right) \left( \frac{H n_0 E_1}{2 V_T} x + \frac{M_1}{2} \right) \notag\\
& \qquad +\frac{E_1}{V_T} \left( (1+Z) e M_{21} - \left( \frac{1-Z^2}{1+Z} \right) \frac{P_{2}^{(+)}}{e l_s} e^{x/l_s} + \left( \frac{1-Z^2}{1+Z} \right) \frac{P_{2}^{(-)}}{e l_s} e^{-x/l_s}  \right) \notag\\
& \qquad + (1+Z)\frac{H n_0 e E_1}{V_T^2} \left(F_2 -  \frac{H E_1^2}{V_T} x  + \frac{ l_s P_{2}^{(+)}}{\epsilon} e^{x/l_s} - \frac{ l_s P_{2}^{(-)}}{\epsilon} e^{-x/l_s} \right) \notag\\
& \qquad + (1+Z)\frac{e}{V_T} \left( \frac{H n_0 E_1}{V_T} x + M_1  \right) \left( -  \frac{H E_1^2}{V_T} + \frac{ P_{2}^{(+)}}{\epsilon} e^{x/l_s} + \frac{ P_{2}^{(-)}}{\epsilon} e^{-x/l_s}   \right). 
\label{eq:PartialDeltaN3}
\end{align}
\end{widetext}
The solution to this second order differential equation, with two new integration constants $\beta_1^{(+)}$ and $\beta_1^{(-)}$, is
\begin{align}
\rho_3 = & \alpha_1 +  \alpha_2 x +  \beta_{1}^{(+)} e^{x/l_s} +  \beta_{1}^{(-)} e^{-x/l_s} +  \beta_{2}^{(+)} x e^{x/l_s} \notag\\
 & +  \beta_{2}^{(-)} x e^{-x/l_s} +  \beta_{3}^{(+)} x^2 e^{x/l_s} + e \beta_{3}^{(-)} x^2 e^{-x/l_s},
\label{eq:DeltaN3}
\end{align}
where, with substitution of $F_2$ from \eqref{eq:F2Solved},
\begin{align}
& \alpha_1 = - \left( \frac{D_b - D_a}{D_a D_b} \right)\frac{M_1  e l_s^2}{2} - 2\frac{\epsilon E_1 M_{21} }{n_0}  + 2\frac{\epsilon H M_1 E_1^2}{n_0 V_T},\label{eq:alpha1}\\
& \alpha_2 =  -\left( \frac{D_b - D_a}{D_a D_b} \right)\frac{\epsilon H  E_1 }{2 (1+Z)}  + 2 \frac{\epsilon H^2 E_1^3}{V_T^2},\label{eq:alpha2}\\
& \beta_{2}^{(+)} = \left(- \left( \frac{1-Z^2}{1+Z} \right) \frac{E_1}{2 V_T} + \frac{H E_1}{4 V_T}  + \frac{M_1}{2 n_0 l_s}\right)   P_2^{(+)}\\
& \beta_{2}^{(-)} = \left(- \left( \frac{1-Z^2}{1+Z} \right) \frac{E_1}{2 V_T} + \frac{H E_1}{4 V_T}  - \frac{M_1}{2 n_0 l_s}\right)   P_2^{(-)}\\
& \beta_{3}^{(+)} = \frac{H  E_1}{4 V_T l_s}  P_{2}^{(+)}, \qquad
 \beta_{3}^{(-)} = - \frac{H  E_1}{4 V_T l_s}  P_2^{(-)}.
\end{align}
From \eqref{eq:DeltaN3} we infer that $N_{a3}$ and $Z N_{b3}$ may include polynomials whose terms that are quadratic and higher are equal, and may include exponential terms that differ.

Substituting $\rho_3$ from \eqref{eq:DeltaN3} into Gauss's Law \eqref{eq:GaussTerms}, and integrating $\partial_x E_3$ gives
\begin{align}
&E_3 =  F_3 + \frac{ \alpha_1}{\epsilon} x + \frac{ \alpha_2}{2\epsilon} x^2 \notag\\
&\quad + \frac{ l_s}{\epsilon} \left( \beta_{1}^{(+)} - l_s \beta_{2}^{(+)} + 2 l_s^2 \beta_{3}^{(+)} \right) e^{x/l_s} \notag\\
&\quad - \frac{ l_s}{\epsilon} \left( \beta_{1}^{(-)} + l_s \beta_{2}^{(-)} + 2 l_s^2 \beta_{3}^{(-)} \right) e^{-x/l_s} \notag\\
&\quad + \frac{ l_s}{\epsilon} \left( \beta_{2}^{(+)} - 2 l_s \beta_{3}^{(+)} \right) x e^{x/l_s} \notag\\
&\quad - \frac{ l_s}{\epsilon}\left( \beta_{2}^{(-)} + 2 l_s \beta_{3}^{(-)} \right) x e^{-x/l_s} \notag\\
&\quad + \frac{ l_s}{\epsilon}\beta_{3}^{(+)} x^2 e^{x/l_s} - \frac{ l_s}{\epsilon}\beta_{3}^{(-)} x^2 e^{-x/l_s},
\label{eq:E3}
\end{align}
where $F_3$ is a new integration constant, with units V-s$^{3/2}$/m.

Substitution of the coefficients $N_{a1}$ from \eqref{eq:Na1Nb1Solved}, $N_{a2}$ from \eqref{eq:Na2Generic}, $E_2$ from \eqref{eq:E2Solved}, $J_{a3}$ from \eqref{eq:Ja3}, and $E_3$ from \eqref{eq:E3} into the $n=3$ Planck-Nernst equation for electrons \eqref{eq:PNa3} yields
\begin{widetext}
\begin{align}
&\partial_x N_{a3} = -\left(\frac{M_{20} E_1}{V_T} + \frac{M_1 F_2}{V_T} + \frac{n_0 F_3}{V_T} + \frac{K_{3}}{D_a} \right) \notag\\
& \quad + \left(\frac{H M_1 E_1^2}{V_T^2} - \frac{M_{21} E_1}{V_T} - \frac{H n_0 F_2 E_1}{V_T^2} - \frac{ n_0 \alpha_1}{\epsilon V_T} - \frac{M_1}{2 D_a} \right)x \notag\\
& \quad + \left( \frac{H^2 n_0 E_1^3}{V_T^3}  - \frac{ n_0 \alpha_2}{2 \epsilon V_T} - \frac{H n_0 E_1}{4 D_a V_T}  \right)x^2 \notag\\
& \quad + \left( \frac{E_1 P_{2}^{(+)}}{(1+Z)e V_T} - \frac{ l_s M_1 P_{2}^{(+)}}{\epsilon V_T} - \frac{ n_0 l_s}{\epsilon V_T} \left( \beta_{1}^{(+)} - l_s \beta_{2}^{(+)} + 2 l_s^2 \beta_{3}^{(+)} \right) \right) e^{x/l_s} \notag\\
& \quad + \left( \frac{E_1 P_{2}^{(-)}}{(1+Z)e V_T} + \frac{ l_s M_1 P_{2}^{(-)}}{\epsilon V_T} + \frac{ n_0 l_s}{\epsilon V_T} \left( \beta_{1}^{(-)} + l_s \beta_{2}^{(-)} + 2 l_s^2 \beta_{3}^{(-)} \right)  \right) e^{-x/l_s} \notag\\
& \quad + \left( -\frac{ n_0 l_s H E_1 P_{2}^{(+)}}{\epsilon V_T^2} - \frac{ n_0 l_s}{\epsilon V_T} \left( \beta_{2}^{(+)} - 2 l_s \beta_{3}^{(+)}\right) \right) x e^{x/l_s} \notag\\
& \quad + \left(\frac{ n_0 l_s H E_1 P_{2}^{(-)}}{\epsilon V_T^2} + \frac{ n_0 l_s}{\epsilon V_T} \left( \beta_{2}^{(-)} + 2 l_s \beta_{3}^{(-)}\right)\right) x e^{-x/l_s} \notag\\
& \quad - \frac{ n_0 l_s}{\epsilon V_T} \beta_{3}^{(+)} x^2 e^{x/l_s}+ \frac{ n_0 l_s}{\epsilon V_T} \beta_{3}^{(-)} x^2 e^{-x/l_s}.
\end{align}
\end{widetext}
The solution to this first order differential equation, with one new integration constant $M_{30}$, is
\begin{align}
&N_{a3} = M_{30} + M_{31}x + M_{32}x^2 + M_{33}x^3 \notag\\
&\quad + \gamma_{1}^{(+)} e^{x/l_s} + \gamma_{1}^{(-)} e^{-x/l_s} + \gamma_{2}^{(+)}x e^{x/l_s} \notag\\
& \quad  + \gamma_{2}^{(-)} x e^{-x/l_s} + \gamma_{3}^{(+)} x^2 e^{x/l_s} + \gamma_{3}^{(-)} x^2 e^{-x/l_s},
\label{eq:Na3}
\end{align}
where 
\begin{align}
&M_{31} = - \left( \frac{M_{20} E_1}{V_T} + \frac{M_1 F_2}{V_T} + \frac{n_0 F_3}{V_T} + \frac{K_{3}}{D_a} \right),
\label{eq:M31}\\
&M_{32} = \frac{1}{2} \left( \frac{H M_1 E_1^2}{V_T^2} - \frac{M_{21} E_1}{V_T} \right. \notag\\
&\qquad \qquad \qquad \left.- \frac{H n_0 F_2 E_1}{V_T^2} - \frac{ n_0 \alpha_1}{\epsilon V_T} - \frac{M_1}{2 D_a}  \right),\\
&M_{33} = \frac{1}{3} \left( \frac{H^2 n_0 E_1^3}{V_T^3}  - \frac{ n_0 \alpha_2}{2 \epsilon V_T} - \frac{H n_0 E_1}{4 D_a V_T} \right),\\
&\gamma_{1}^{(+)} = \left(\frac{(H+1) l_s E_1}{V_T}  - \frac{M_1}{n_0} \right)\frac{ P_{2}^{(+)} }{(1+Z)e}\notag\\
&\qquad \quad - \frac{1}{(1+Z)e}\left( \beta_{1}^{(+)} - 2 l_s \beta_{2}^{(+)} + 6 l_s^2 \beta_{3}^{(+)} \right) ,\\
&\gamma_{1}^{(-)} =  -\left( \frac{M_1}{n_0} + \frac{(H+1) l_s E_1}{V_T} \right) \frac{P_{2}^{(-)} }{(1+Z)e}\notag\\
&\qquad \quad - \frac{1}{(1+Z)e}\left( \beta_{1}^{(-)} + 2 l_s \beta_{2}^{(-)} + 6 l_s^2 \beta_{3}^{(-)} \right) ,\\
&\gamma_{2}^{(+)} = -\frac{H E_1 P_{2}^{(+)}}{(1+Z)e V_T} - \frac{1}{(1+Z)e} \left(\beta_{2}^{(+)} - 4 l_s \beta_{3}^{(+)} \right),\\
&\gamma_{2}^{(-)} = -\frac{H E_1 P_{2}^{(-)}}{(1+Z)e V_T} - \frac{1}{(1+Z)e} \left(\beta_{2}^{(-)} + 4 l_s \beta_{3}^{(-)} \right),\\
&\gamma_{3}^{(+)} =  -\frac{\beta_{3}^{(+)}}{(1+Z)e} ,\qquad \gamma_{3}^{(-)} =  -\frac{\beta_{3}^{(-)}}{(1+Z)e}.
\end{align}
Use of \eqref{eq:rho_n-J_n} for $n=3$ ($\rho_3 = e(ZN_{b3}-N_{a3}) $), $\rho_3$ from \eqref{eq:DeltaN3}, and $N_{a3}$ from \eqref{eq:Na3} gives
\begin{align}
&N_{b3} = \left(\frac{\alpha_1  + e M_{30}}{Z e}\right) + \left(\frac{\alpha_2  + e M_{31} }{Z e} \right) x \notag\\
&\quad + \left(\frac{M_{32}}{Z} \right) x^2 + \left(\frac{M_{33}}{Z} \right) x^3 + \left( \frac{\beta_{1}^{(+)}  + e \gamma_{1}^{(+)}}{Z e} \right)e^{x/l_s} \notag\\
&\quad + \left( \frac{\beta_{1}^{(-)}  + e \gamma_{1}^{(-)}}{Z e} \right)e^{-x/l_s} + \left( \frac{\beta_{2}^{(+)}  + e \gamma_{2}^{(+)}}{Z e} \right)x e^{x/l_s} \notag\\
&\quad + \left( \frac{\beta_{2}^{(-)} + e \gamma_{2}^{(-)}}{Z e} \right)x e^{-x/l_s} + \left( \frac{\beta_{3}^{(+)}  + e \gamma_{3}^{(+)}}{Z e} \right)x^2 e^{x/l_s} \notag\\
&\quad  + \left( \frac{\beta_{3}^{(-)}  + e \gamma_{3}^{(-)}}{Z e} \right)x^2 e^{-x/l_s}.
\end{align}

%
We now have $J_{a3}$, $J_{b3}$, $E_3$, $N_{a3}$ and $N_{b3}$ with five constants of integration, $K_{3}$, $\beta_{1}^{(+)}$, $\beta_{1}^{(-)}$, $F_3$, and $M_{30}$.  To reduce this from five to four, 
we add \eqref{eq:PNa3} divided by $D_a$ and \eqref{eq:PNb3} divided by $D_b$, 
\begin{equation}
\begin{split}
\frac{J_{a3}}{D_a} + \frac{J_{b3}}{D_b} + \partial_x(N_{a3} + N_{b3}) = \frac{1}{V_T} \left(\frac{\rho_2 E_1}{e} + \frac{\rho_1 E_2}{e}  \right). 
\label{eq:SumPNa3PNb3}
\end{split}
\end{equation}
As for $n=2$, 
in \eqref{eq:SumPNa3PNb3} the exponential, quadratic, and linear terms match, by construction.  A new constraint, however, is found by comparing the constant terms,
\begin{equation}
\begin{split}
&\frac{K_3}{D_a} + \frac{K_3}{Z D_b} + \frac{\epsilon E_1}{2 Z e D_b} + M_{31} + \frac{\alpha_2 + e M_{31}}{Z e} =  -  \frac{H \epsilon E_1^3}{V_T^2 e} .
\end{split}
\end{equation}
Substituting $M_{31}$ from \eqref{eq:M31} yields
\begin{align}
&F_3 = \left(\frac{D_b - D_a}{(1+Z)D_a D_b}\right) K_3 - \frac{M_{20} E_1}{n_0} - \frac{M_1 F_2}{n_0} \notag\\
&\qquad - \left( \frac{1}{(1+Z)e} \right) \left( \frac{Z H \epsilon E_1^3}{V_T^2} + \frac{\epsilon E_1}{2  D_b} + \alpha_2 \right),
\end{align}
a relation between integration constants $F_3$ and $K_3$.  

We are thus left with four independent constants of integration, which we take to be $K_3$, $M_{30}$, $\beta_1^{(+)}$, and $\beta_1^{(-)}$. 
The reaction rates for electrons and ions at each interface, needed to produce the correct ion fluxes, provide the four conditions necessary to solve for these constants.

For completeness, we note that from Gauss's Law at each surface \eqref{eq:GaussSurfaceTerms},
\begin{equation}
E_3(0) = \frac{\Sigma_3^{(0)}}{\epsilon}, \quad E_3(L) = -\frac{\Sigma_3^{(L)}}{\epsilon}.
\end{equation}
Then, \eqref{eq:E3} yields
\begin{align}
&\Sigma_3^{(0)} = \epsilon F_3 +  l_s \left( \beta_{1}^{(+)} - l_s \beta_{2}^{(+)} + 2 l_s^2 \beta_{3}^{(+)} \right)  \notag\\
&\quad -  l_s \left( \beta_{1}^{(-)} + l_s \beta_{2}^{(-)} + 2 l_s^2 \beta_{3}^{(-)} \right), \\
&\Sigma_3^{(L)} = - \epsilon F_3 -  \alpha_1 L - \frac{ \alpha_2}{2} L^2 \notag\\
&\quad -  l_s \left( \beta_{1}^{(+)} - l_s \beta_{2}^{(+)} + 2 l_s^2 \beta_{3}^{(+)} \right) e^{L/l_s} \notag\\
&\quad +  l_s \left( \beta_{1}^{(-)} + l_s \beta_{2}^{(-)} + 2 l_s^2 \beta_{3}^{(-)} \right) e^{-L/l_s} \notag\\
&\quad -  l_s \left( \beta_{2}^{(+)} - 2 l_s \beta_{3}^{(+)} \right) L e^{L/l_s} \notag\\
&\quad +  l_s \left( \beta_{2}^{(-)} + 2 l_s \beta_{3}^{(-)} \right) L e^{-L/l_s} \notag\\
&\quad -  l_s\beta_{3}^{(+)} L^2 e^{L/l_s} +  l_s\beta_{3}^{(-)} L^2 e^{-L/l_s}.
\end{align}
We have verified that
\begin{equation}
\Sigma_3^{(0)}+\Sigma_3^{(L)} +  \int_0^L \rho_3 dx =0,
\end{equation}
so there is no net charge in the system.

\bibliography{OxidationBib}{}
\bibliographystyle{unsrt}

\end{document}